# Ultrafast coherent interlayer phonon dynamics in atomically thin layers of MnBi$_2$Te$_4$

F. Michael Bartram[1,2], Yu-Chen Leng[3], Yongchao Wang[4], Liangyang Liu[1], Xue Chen[3], Huining Peng[1], Hao Li[5], Pu Yu[1,5,6], Yang Wu[7,8], Miao-Ling Lin[3], Jinsong Zhang[1,5], Ping-Heng Tan[3] and Luyi Yang[1,5,*]

[1]State Key Laboratory of Low Dimensional Quantum Physics, Department of Physics, Tsinghua University, Beijing 100084, China

[2]Department of physics, University of Toronto, Toronto, Ontario M5S 1A7, Canada

[3]State Key Laboratory of Superlattices and Microstructures, Institute of Semiconductors, Chinese Academy of Sciences, Beijing 100083, China

[4]Beijing Innovation Center for Future Chips, Tsinghua University, Beijing, China

[5]Frontier Science Center for Quantum Information, Beijing 100084, China

[6]RIKEN Center for Emergent Matter Science (CEMS), Wako 351-198, Japan

[7]Tsinghua-Foxconn Nanotechnology Research Center, Department of Physics, Tsinghua University, Beijing 100084, China

[8]Beijing Univ Chem Technol, Coll Sci, Beijing, 100029, China

*Email: luyi-yang@mail.tsinghua.edu.cn

**Abstract**

The atomically thin MnBi$_2$Te$_4$ crystal is a novel magnetic topological insulator, exhibiting exotic quantum physics. Here we report a systematic investigation of ultrafast carrier dynamics and coherent interlayer phonons in few-layer MnBi$_2$Te$_4$ as a function of layer number using time-resolved pump-probe reflectivity spectroscopy. Pronounced coherent phonon oscillations from the interlayer breathing mode are directly observed in the time domain. We find that the coherent oscillation frequency, the photocarrier and coherent phonon decay rates all depend sensitively on the sample thickness. The time-resolved measurements are complemented by ultralow-frequency Raman spectroscopy measurements, which both confirm the interlayer breathing mode and additionally enable observation of the interlayer shear mode. The layer dependence of these modes allows us to extract both the out-of-plane and in-plane interlayer force constants. Our studies not only reveal the interlayer van der Waals coupling strengths, but also shed light on the ultrafast optical properties of this novel two-dimensional material.

**Introduction**

The marriage between topology and magnetism can give birth to many exotic quantum phases and phenomena such as quantum anomalous Hall effect [1-4] and axion electrodynamics [5-8]. Recently, few-layer $MnBi_2Te_4$ (MBT) crystals have emerged as a new platform for exploring these phases [9-33]. MBT is a new class of two-dimensional (2D) topological magnetic material. The crystal structure of MBT is formed by Te-Bi-Te-Mn-Te-Bi-Te septuple layers (SLs) stacked on top of each other along the *c* direction (Fig. 1a). The topological properties are inherited from the *p* orbitals of Bi and Te atoms, while the magnetism originates from the *d* bands of Mn atoms. The adjacent SLs are bonded by van der Waals (vdW) forces and can be cleaved to thin layers. Both electronic and magnetic properties depend sensitively on the sample thickness. Transport studies have revealed rich quantized transport phenomena in few-layer MBT (5-9 SL), suggesting odd- and even- layer MBT samples may host the quantum anomalous Hall and axion insulator states respectively [24-27]. Magnetic circular dichroism measurements have revealed the layer-dependent magnetism with an odd-even layer-number effect, verifying the A-type antiferromagnetic structure in few-layer MBT [28,29].

In 2D vdW materials, lattice vibrations, especially interlayer phonon modes, have not only provided unique capabilities in determining layer thickness, stacking order, and interface coupling strength [34-36], but have also played an important role in engineering novel electronic, thermal and magnetic properties in 2D vdW homo/heterostructures [37-42]. For instance, in twisted bilayer graphene interlayer conductance and thermal transport are mediated by the interlayer phonons [39-42]. Several Raman studies have been performed recently, primarily on high-frequency intralayer phonons which have demonstrated spin-phonon coupling [31-33]. In the low-frequency regime, magnon modes have been observed [43], and a systematic study of interlayer breathing mode phonons showed a very large interfacial coupling which produced somewhat unusual results [44]. These interlayer phonons provide information about vdW coupling strength and electron-phonon scattering mechanisms, and a good understanding of their properties is useful for future work on MBT, for example in manipulating magnetic and topological states with light at a fast speed [23,45,46], in the design and probe of heterostructure devices [37,38], or simply as a method for determining sample thicknesses [35,36].

Here we report the first ultrafast optical pump-probe reflectivity measurements in few-layer MBT, accompanied by ultralow frequency Raman data, of carrier and coherent interlayer phonon dynamics as a function of sample thickness, with the layer number varying from 4 to 25. We observe pronounced oscillatory signals due to coherent interlayer breathing mode phonons in transient reflection measurements and find that physical properties such as the oscillation frequency, its decay time, and the carrier decay rate all depend strongly on the sample thickness. The frequency is particularly consistent and easily measured, which provides a convenient, contact-free and non-destructive tool to characterize sample thickness, especially for ultrafast studies. The increasing decay rates for thinner samples suggest that the surface and interface have a strong impact on the phonon scattering and photocarrier relaxation. The time-domain measurements are complemented by ultralow-frequency (<10 cm$^{-1}$) Raman studies. In addition to the interlayer breathing modes, polarized Raman measurements enable the observation of the much weaker interlayer shear modes for the first time. The measured phonon frequencies all show a typical blueshift with decreasing layer number and can be fit using a linear chain model, from which the out-of-plane and in-plane interlayer force constants are calculated to be $(5.7 \pm 0.1) * 10^{19}$ N/m³ and $(2.9 \pm 0.1) * 10^{19}$ N/m³ respectively. The force constants allow us to calculate other mechanical parameters such as sound velocities, elastic constants and acoustic impedance. These measurements not only uncover the interlayer coupling strengths but also provide crucial information on coherent phonon and carrier lifetimes and relaxation mechanisms.

## Results

### Atomic force microscopy measurements

Few-layer MBT crystals were prepared by mechanical exfoliation as described previously [25]. Figure 1b shows an optical image of one of the sample regions used in this work. The layer thickness was determined by atomic force microscopy (AFM) measurements (Fig. 1c-e), where clear step sizes of around 1.4 nm can be observed, consistent with previous studies [24-26,29].

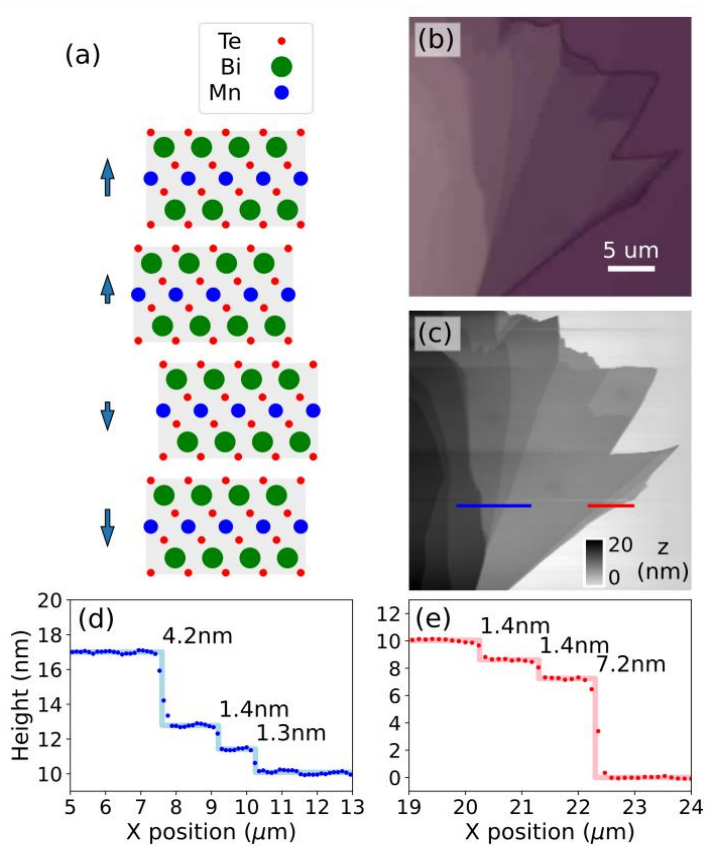

*Figure 1 Atomic force microscopy measurements. A simple diagram (a) of the atomic structure of MBT and the motion induced by the lowest-frequency interlayer breathing phonons for a 4-layer region of the sample. Optical (b) and AFM (c) images of a region of one of our exfoliated samples, along with cuts (d,e) showing various steps, which match well with the known ~1.4 nm layer thickness.*

**Time-resolved reflectivity measurements**

We measured the transient reflectivity with a standard two-color pump-probe setup, the results of which can be seen in Figure 2. In the raw data, shown in Fig. 2a, we clearly see the presence of three distinct components. First, there is a large but short-lived change which lasts only 2-3 picoseconds with negligible layer dependence before vanishing, associated with hot electrons which rapidly transfer energy to the surrounding lattice via electron-electron and electron-phonon interactions. Next, we see a simple exponential decay, with a lifetime of around 100 ps depending on the layer thickness, probably reflecting photocarrier scattering and electron-hole recombination processes. The final component is an oscillatory signal which decays over approximately 30-50 picoseconds, and then briefly reappears at t ≈ 100 ps (see inset) before decaying away once more. Note that the oscillatory signal is very strong, about 20% of the total transient reflection. To show the oscillatory signal more clearly, the data after the first two components are subtracted is plotted again in Fig. 2b, along with the FFT of the data in Fig 2c, where you can see a clear upward shift in frequency as the sample thickness decreases from ~130 GHz in 9 SL to ~180 GHz in 7 SL. The inset shows the FFT of the echoes, which have frequencies that are identical to the main signals.

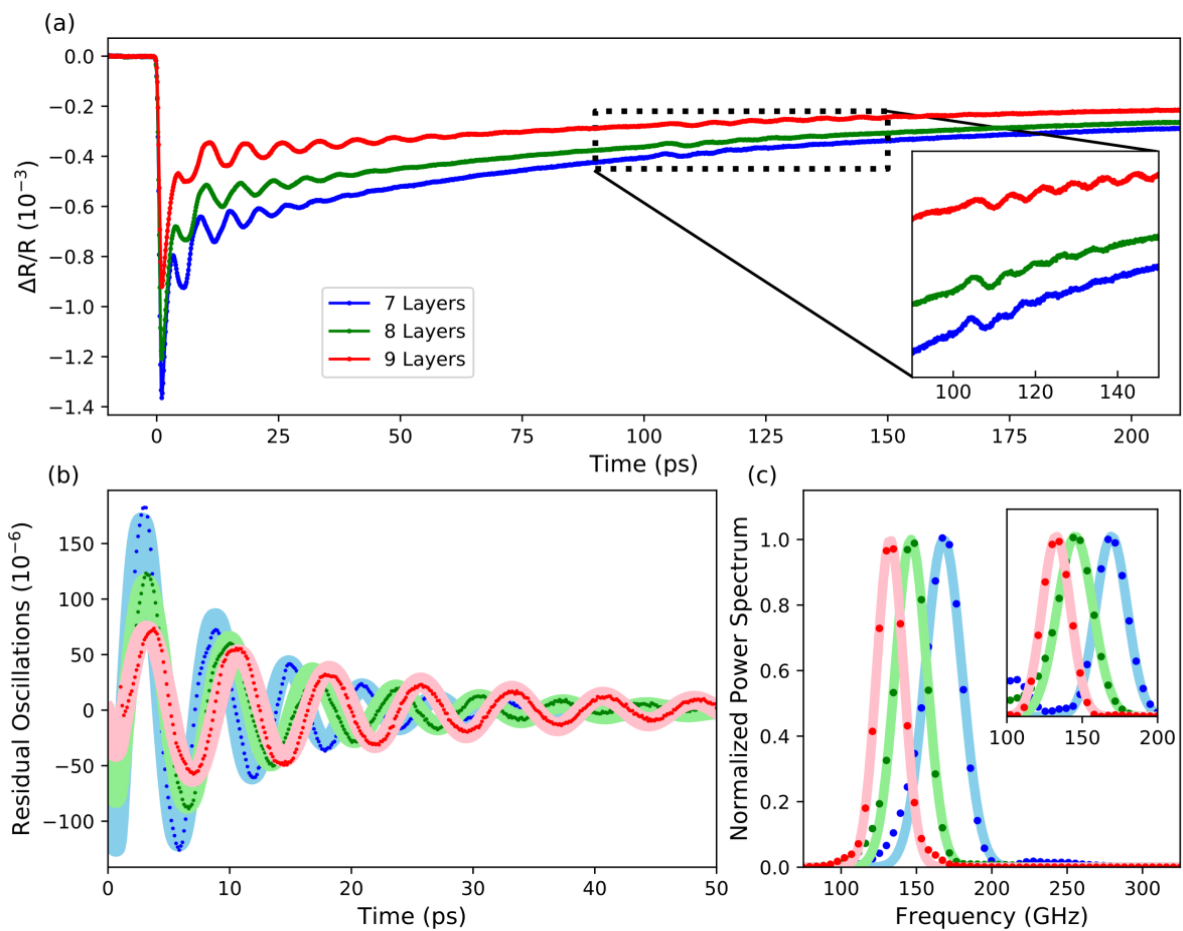

*Figure 2 Time-resolved reflectivity data measured at various thicknesses of exfoliated MnBi$_2$Te$_4$. The raw data (a) for three different regions shows the presence of a very short-lived spike, an exponential decay, and a phonon oscillation, with the inset highlighting an echo of the oscillation which occurs approximately 100 ps after the pump pulse. Subtracting the first two components allows the phonon oscillation to be seen more clearly (b), where solid lines show fits to an exponentially decaying sinusoid, which match well aside from slight discrepancies at very short times*

*(likely due to issues arising from subtraction of the very short-lived component). Plotting the data after FFT (c) highlights the clear shift in frequency between the different thicknesses. The inset in (c) shows the FFT of the echoes at ~100 ps, which have identical layer dependence.*

In a pump-probe experiment, coherent phonons created by the pump pulse can modulate the optical dielectric function and can therefore be detected via reflectivity changes measured by the synchronized probe pulse. These phonons can be created via rapid expansions or contractions induced by, for example, temperature changes [47,48] or carrier excitation [49], or through a stimulated Raman scattering process [50,51]. Because the observed oscillation frequencies are much smaller than the intralayer vibration frequencies measured by Raman spectroscopy [31] (also see discussions below), we attribute these results to coherent interlayer phonons, which for ultrathin samples will form standing waves with strongly layer-dependent frequencies. In our measurements, the polarization of the pump beam does not affect the observed signal (see Supplementary Information Note 3). This suggests that the interlayer vibrations are in the out-of-plane direction (the breathing mode), as in-plane vibrations (the shear mode) would show a phase shift when the pump polarization is changed [51,52], a fact which has been utilized to detect very weak coherent interlayer shear mode phonons in graphene [52,53]. The assignment of the coherent phonon mode is further confirmed by our polarization-resolved ultralow-frequency Raman spectroscopy measurements as discussed below.

This mechanism also explains the echo of the signal at 100 ps, as coupling between the sample and the ~300 nm thick silica layer on the substrate means that acoustic phonons can be generated, travel away from the surface, reflect from the internal silica/Si interface, and return to the surface, where they then re-excite the interlayer phonon in the sample. The timing of the reappearance gives a velocity of about 6 km/s, which matches well with the speed of longitudinal acoustic phonons in silica [54]. The echo signals indicate that few layer MBT can be used as an optical transducer to launch coherent acoustic phonons in $SiO_2$.

In Figure 3a, we plot the extracted oscillation frequency at a variety of sample locations for a given layer thickness as a function of the inverse layer number (1/*N*). The frequency decreases with increasing sample thickness from ~300 GHz in 4 SL to ~50 GHz in 25 SL samples.

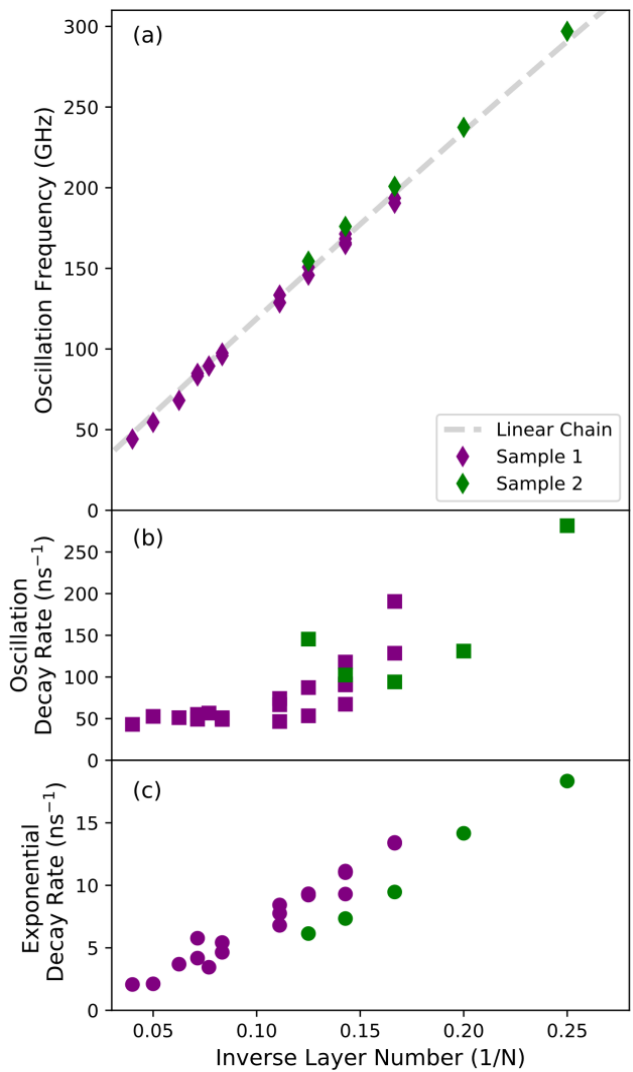

*Figure 3 Thickness dependence of various parameters of the time-resolved reflectivity data. The oscillation frequency (a) increases as the sample becomes thinner and fits well to a linear-chain model $f(N) = f_{0,B} \sin\left(\frac{\pi}{2N}\right)$ with $f_{0,B} = 760$ GHz. The decay rates of both the oscillatory component (b) and the simple exponentially decaying component (c) also increase for thinner samples, with the latter being particularly noticeable. Note that statistical uncertainties from fitting are very small (≈0.1%, 1% and 0.5% for points in (a), (b) and (c) respectively, all within marker sizes), but practical uncertainties are clearly much larger due to factors such as variation across sample locations.*

We analyze how the frequency of these phonons vary with thickness using a basic freestanding linear chain model [34-36], in which each SL is moving as a unit and only the nearest layer interaction is considered (see Supplementary Information Note 9). A similar model has been used to explain the interlayer vibration modes in a variety of other 2D layered materials [34-36,55-60]. An *N*-layer system hosts *N*-1 breathing/shear modes with frequencies $f(N,n) = f_0 \sin(n\pi/(2N))$, where $n = 1, 2, \cdots, N-1$ is the branch index and $f_0 = (1/\pi)\sqrt{k/\mu}$, where $k = k_z$ (or $k_x$) is the out-of-plane (or in-plane) force constant per unit area between the adjacent layers and $\mu$ is the mass per unit area. Fitting our results to $f(N,1)$ gives a resulting fit parameter $f_{0,B} = 760 \pm 5$ GHz for the breathing mode, which we plot as a dashed grey line in Figure 3a, showing excellent agreement of this theoretical model with the experimental data. Note that when the number of layers $N \to \infty$, these modes correspond to longitudinal acoustic phonons along the out-of-plane direction with velocity $v = d\sqrt{k/\mu}$, with $d$ the spacing between layers. Therefore, the longitudinal sound velocity is given by $v_B = \pi d f_0 = 3.25$ km/s. This value is comparable to $v_B$ in other 2D layered materials such as $MoS_2$ (see Table S1 in Supplementary Information Note 8).

One obvious advantage of time-resolved experiments is that we can directly measure the phonon and photocarrier lifetimes. We plot the decay rate of both the oscillatory component and the simple exponential decay in Fig. 3b and 3c respectively. These decay rates noticeably increase for the thinner regions of the sample. For a given thickness, we performed multiple measurements on different locations. Both the phonon decay and carrier relaxation are very sensitive to the local environment especially for the thinner samples shown as scattered data points for the same thickness in Fig. 3b and 3c, suggesting that phonon and carrier relaxation is influenced strongly by the surface and interface (e.g., surface roughness, adatoms and charges trapped at the interface, which all act as scatterers). On the other hand, the coherent phonon frequency is measured to be robust against locations (Fig. 3a).

In general, coherent phonon dephasing in solids is governed by scattering at boundaries, lattice defects, carriers or population decaying into lower energy acoustic phonons via anharmonic interactions. As shown in Fig. 3b, for thicker samples (> 10 layers) the phonon decay rate is ~50 GHz and is independent of the thickness and sample locations. By contrast, in the thin layer regime (< 10 layers) the decay rate grows rapidly with decreasing thickness and depends strongly on the location, indicating that the surface and interface play a dominant role. As demonstrated by the echo of the signal, the interlayer phonons can couple to acoustic phonons in the substrate, which may further increase the interfacial contribution to the decay rate. Furthermore, the layer dependence of phonon decay rate can also result from stronger electron-phonon coupling in thinner samples than in thicker ones [31], similar to topological materials such as $Bi_2Se_3$ and $WP_2$ [61,62]. Additional insight into the anharmonic contribution to the phonon decay rate can be obtained from temperature-dependent data (Fig. S5 in Supplementary Information Note 5). If the anharmonic contribution were large, the decay rate (linewidth) would increase significantly upon increasing temperature [63]. Our measurement shows that the decay rate stays roughly constant with temperature from 3 K to room temperature, suggesting that the anharmonic interaction is not important. We have found that the transient reflectivity simply scales linearly with the pump fluence (see Supplementary Information Note 6), which eliminates scattering with photoinduced carriers as a prominent dephasing factor.

The photoinduced carrier relaxation is determined by carrier scatterings with defects and lattice and electron-hole recombination, etc. The increasing decay rate in thinner samples from ~2 GHz in 20 SL to ~18 GHz in 4 SL shown in Fig. 3c indicates accelerated scattering and recombination rates at the surface and interface compared to the inner layers. The approximate linear scaling of the decay rate with ($1/N$) can be understood with a simple model, in which the total decay rate is a weighted average between the inner layer decay rate and surface/interface decay rate. Similar behavior has been observed in 2D $MoS_2$ and $Bi_2Te_3$ [64,65].

**Raman spectroscopy measurements**

To confirm our observations, we also performed polarized Raman measurements on our sample, shown in Fig. 4a and 4b. The 6 high-frequency peaks (at ~27, 48, 68, 105, 115 and 141 $cm^{-1}$) have been observed in previous studies [31] and are intralayer phonon modes that are Raman active both in bulk and few-layer samples. We assign these modes corresponding to their symmetry class [22] based on their polarization dependence (see Supplementary Information Note 1 for more details). The out-of-plane vibrations ($A_{1g}$ modes) can be observed only in a parallel-polarized geometry, whereas the in-plane vibrations ($E_g$ modes) are detectable in both parallel- and cross-polarized geometries.

At low frequencies, we also observe new peaks which we attribute to the interlayer phonons. In addition to the breathing mode seen in the time-resolved data, (labeled $B^{(1)}$), Raman measurements also enable observation of the much weaker shear mode (labeled $S^{(1)}$) for the first time. Note that unlike the breathing modes, the shear modes are sensitive to applied in-plane strain [66], which potentially can be used as a strain probe in MBT and heterostructures when engineering electronic and magnetic properties with strain. The $B^{(1)}$ peak intensity is remarkably strong compared to all other peaks, which is consistent with the pronounced coherent oscillations in the time-resolved measurements as the coherent phonons are generated via the stimulated Raman scattering process. The frequencies of the $B^{(1)}$ and $S^{(1)}$ modes are plotted in Fig. 4c alongside the time-resolved data, showing the excellent agreement of the $B^{(1)}$ frequencies with those observed in the transient reflectivity. In addition, we show that the $S^{(1)}$ modes also fit well to the linear chain model, with a slightly lower $f_{0,S} = 545 \pm 7$ GHz. The linewidths of these modes (plotted in the Supplementary Information) are very noisy, making any layer dependence difficult to observe, but are reasonable compared to the decay rates in the time-resolved data. While some small additional peaks can be observed in the parallel-polarized data (e.g. around ~20 cm$^{-1}$ in Fig. 4a), they have been left intentionally unlabeled as their precise origin is unclear and is not the focus of this work. Note that only the $B^{(1)}$ mode is seen in the time-resolved measurements, but since Raman scattering intensity is correlated with the amplitude of coherent phonons that can be generated [51], these other modes may simply be too weak to be observed.

Since both the $B^{(1)}$ and $S^{(1)}$ modes show a large blue shift from ~ 120 GHz in 10 SL to ~300 GHz in 4 SL and from ~100 GHz in 9 SL to 200 GHz in 4 SL, respectively, the frequency dispersions of these interlayer modes could be useful for determining the thickness of few-layer MBT. By contrast, among the high frequency intralayer modes, only the $A_{1g}^{(1)}$ mode has a noticeable change [57], red-shifting from ~47.5 to ~44.5 cm$^{-1}$ (a difference of ~90 GHz) from 10 SL to 4 SL (see Supplementary Information), similar to previously reported observations [31]. We also point out that while in Raman spectroscopy these interlayer mode frequencies are very low and may be difficult to measure, especially for thick samples (where for our setup, above ~13 SL the frequencies drop below the detection limit), time-resolved measurements can be used to quickly and accurately obtain the breathing mode frequency across a wide thickness range. Acquiring time-resolved data could therefore be a particularly effective tool for nondestructive measurements of MBT sample thickness.

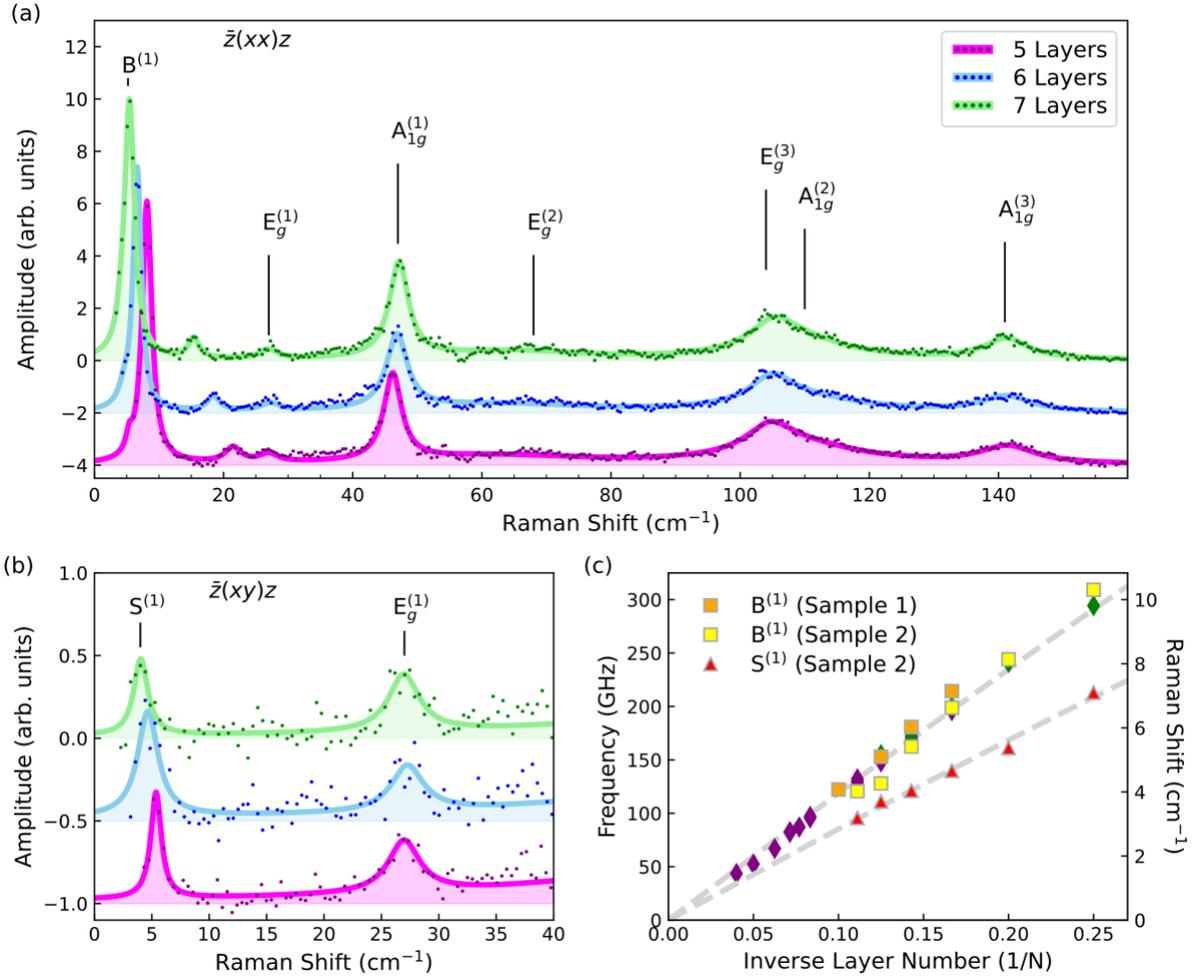

*Figure 4 Polarized Raman measurements on MnBi$_2$Te$_4$. With parallel polarization (a), the data contains 6 peaks at high frequencies which match previously reported Raman data, as well as a very large low-frequency peak corresponding to the phonon breathing mode, labeled B$^{(1)}$. The Rayleigh scattering background is subtracted. In the cross-polarized configuration (b) this large peak vanishes, and an additional peak corresponding to the shear mode, labeled S$^{(1)}$, is revealed. Plotting the frequency of the B$^{(1)}$ and S$^{(1)}$ peaks as a function of thickness (c) along with the previously shown time-resolved data (purple and green diamond symbols for Samples 1 and 2 respectively) shows a good agreement of the breathing mode with the time-resolved data as well as a good fit of the shear mode to the linear chain model with $f_{0,S} = 545$ GHz. Although the statistical uncertainties remain*

*small (≈ 0.1% for B peaks and ≈ 1% for S peaks), compared to the time-resolved data the extracted peak locations show more variation, probably due to the influence of the background, which is very large at such low wavenumbers.*

**Discussion**

As mentioned previously, we have found that for the breathing mode, $f_{0,B} = 760 \pm 5$ GHz from the time-resolved data and for the shear mode, $f_{0,S} = 545 \pm 7$ GHz from the S$^{(1)}$ Raman peaks. Using the B$^{(1)}$ Raman peaks to determine the breathing mode frequency results in a slightly higher and more uncertain $f_{0,B}^R = 780 \pm 15$ GHz, likely due to the large Rayleigh scattering background in parallel-polarized Raman measurements at such low frequencies. Using a mass per unit area per layer for MBT of $\mu = 10.0 * 10^{-6}$ kg/m², we find the out-of-plane and in-plane force constants per unit area to be $k_z = (5.7 \pm 0.1) * 10^{19}$ N/m³ and $k_x = (2.9 \pm 0.1) * 10^{19}$ N/m³. The out-of-plane force constant $k_z$ compares well with a recent Raman study, where they indirectly estimated $k_z = (4.5 \pm 0.7) * 10^{19}$ N/m³ from the Davydov splitting in bulk MnBi$_4$Te$_7$ [32]. Recent first-principles calculations of the interlayer phonon modes in bilayer MBT [23] give force constants of $k_z = 2.85 * 10^{19}$ N/m³ and $k_x = 1.66 * 10^{19}$ N/m³, which like previous studies on Bi$_2$Te$_3$ [59] are much smaller than the experimental values but show a similar ratio between the out-of-plane and in-plane force constants (~1.7 from the theory, compared to ~1.9 from the experiment).

As mentioned in the introduction, a very recent study also measured the interlayer breathing mode phonons using Raman spectroscopy [44]. They observed somewhat unusual results that were attributed to a strong interfacial coupling, supported by the fact that measuring samples on two different substrates gave different results. This fact necessitated the use of a more complicated linear chain model to include the effects of the substrate coupling, which they found explained their results well. Our measurements, in contrast, agree well with a simpler model where interfacial effects are neglected (see Supplementary Information Note 9 for additional discussion). This is somewhat surprising, considering one of the substrates used in their study was the same as ours (SiO$_2$/Si), and suggests that more work should be done to determine under what circumstances these interfacial effects will vary. However, despite this difference, their result for the out-of-plane force constant $k_z = (4.3 - 6.4) * 10^{19}$ N/m³ agrees quite well with ours.

A useful number for comparison with other 2D materials is the per bottom atom rather than per unit area force constants, which are $k_{z/atom} = (9.3 \pm 0.1)$ N/m and $k_{x/atom} = (4.7 \pm 0.1)$ N/m. These per-atom values are smaller than those found in $Bi_2Te_3$, but much larger than graphene (see Table S1 in Supplementary Information Note 8 for details), in line with the relative difficulties of mechanical exfoliation for these different samples. In Table S1, we further compare interlayer phonon frequencies $f_0$, coherent phonon damping times and, acoustic velocities, elastic constants $C_{33}$ and $C_{44}$ and acoustic impedances calculated from the force constants, of MBT with a few representative vdW materials.

We also investigate the interlayer breathing phonons as a function of temperature and magnetic field (see Supplementary Information 5) in order to determine if there are any strong links between the interlayer phonons and magnetic ordering in few-layer MBT. We find that the breathing mode frequency decreases by ~5% from 3 K to room temperature (Fig. S5). This softening results from thermal expansion, which weakens the interlayer coupling with increasing temperature. Note that both magnetic and topological electronic properties are affected by hydrostatic pressure shown by recent theoretical and experimental studies in the MBT family compounds [67-69]; the interlayer phonons provide a sensitive probe of the lattice separation and external strain/pressure. We do not see a noticeable change in the breathing phonon frequency as the temperature crosses the Neel temperature at ~20 K (Fig. S5) nor when a magnetic field of up to 6 T is applied at to the sample 3 K (Fig. S6). Ref. [31] reports that there is a 0.3% frequency shift of the ~48 cm$^{-1}$ $A_{1g}(1)$ mode across the phase transition temperature. We note that as the frequency change due to the spin-phonon coupling is proportional to the phonon frequency itself, similar coupling between interlayer breathing phonons and the magnetic order would likely only produce changes that are too small to observe. For example, in a 6-layer sample with ~200 GHz oscillations, a 0.3% change is only 0.6 GHz, which is within the detection uncertainty.

In conclusion, we have directly observed the interlayer breathing mode in few layer MBT both in the time-domain and in the frequency domain. By measuring the vibration frequency as a function of layer number, we have obtained both the out-of-plane and in-plane interlayer force constants. A good understanding of the typical vdW coupling strength and interlayer vibrational properties will set the stage for engineering electronic and magnetic properties [70] (e.g. vdW heterostructures) as a device performance may be affected through interlayer electron-phonon interactions [37,71-73]. Because the energies of interlayer phonons in MBT are small (~1 meV), we expect that they play an important role in interlayer conductance even at low temperatures [39,41]. Our time-resolved measurements have also revealed that the surface and interface have a strong impact on the relaxation of both the coherent phonons and photocarriers, which may also influence carrier mobility. Therefore, a systematic study of transport properties as a function of layer number is desirable. As proposed in Ref. [23], topological and magnetic phase transitions in MBT can be manipulated via non-linear phonon dynamics using intense terahertz (THz) light pulses. The light pulses induce lattice distortions leading to the separation of the septuple layers, and thus change the magnetic coupling between layers (from AFM to FM) and topological states. This effect should be largest when driving resonantly at the interlayer breathing phonon frequencies measured in this work. A similar approach has been used to realize a topological phase transition in $WTe_2$ [46], in which intense THz pulses drive the interlayer shear mode and induce a structure phase transition from a non-centrosymmetric to centrosymmetric phase, which resulted in the sample changing from a Weyl semimetal to a topologically trivial metal. Our studies also pave the way for future light-driven topological and magnetic orders in few-layer topological magnetic materials [23,45,46], and ultrafast studies of nonequilibrium magnetic dynamics [74] and nonequilibrium axion dynamics [75-78].

## Methods

### Sample Preparation

$MnBi_2Te_4$ single crystals were grown by the direct reaction of $Bi_2Te_3$ and MnTe with a ratio of 1:1 in a vacuum-sealed silica ampoule. The mixture was heated to 973 K and then cooled down to 864 K slowly. The growth method and characterization are similar to those used in previous works [18,25]. The Neel temperature and the critical magnetic field for the spin flop transition of bulk crystals are $T_N$ = 25 K and $H_{c1}$ = 3.6 T (at 1.5 K) respectively [18,25]. These values are shown to be correlated with the Mn content and distribution [79].

MnBi$_2$Te$_4$ flakes were exfoliated onto 285 nm thick SiO$_2$/Si substrates treated by air plasma. The fabrication processes were carried out in an argon-filled glove box. The thickness is determined by the optical contrast and atomic force microscopy (AFM) measurement. For optical measurements, we mounted the sample in a vacuum chamber and immediately pumped down to minimize exposure to ambient environment.

**Time-Resolved Reflectivity Spectroscopy**

Time-resolved measurements were performed using a typical two-color pump-probe setup, where one beam is the direct output of a Ti:Sapphire oscillator (repetition rate: 80 MHz), and the second beam is split off and sent through an OPO to modify the wavelength. After passing through optics and a delay stage, the beams are merged with a dichroic mirror and focused onto the sample using an NA=0.5 aspheric lens, giving a 1-2 μm spot size for the probe beam and a ~5 μm spot for the pump, which is intentionally defocused slightly to allow better overlap with the probe. The reflected light is passed through a filter to eliminate any pump scattering and measured with a balanced photodiode. We modulate the pump beam at 377 kHz with an electro-optical modulator, which allows sensitive lock-in detection of the differential reflectivity. The data in this work was primarily taken using either a 633 nm probe and an 805 nm pump or an 805 nm probe and a 1050 nm pump (see Supplementary Information Note 4 for discussion of other combinations). The probe and pump powers were set to approximately 0.1 and 1 mW (pump fluence ~50 μJ/cm$^2$) average powers respectively to avoid sample damage and laser heating effects. Most measurements were performed at room temperature unless otherwise stated. Temperature and magnetic field dependent measurements were carried out in an optical superconducting magnet system with the magnetic field applied perpendicular to the sample plane.

**Raman Spectroscopy**

The Raman spectroscopy measurements were carried out in backscattering geometry at room temperature using a micro-Raman spectrometer with a charge-coupled device. The laser wavelength was 488 nm and the laser plasma lines were filtered by a reflecting Bragg grating, while the Rayleigh line is suppressed using BragGrate notch filters, achieving a detection limit of ~3 cm$^{-1}$. The sample was mounted in a vacuum chamber. The laser was focused onto the sample via a 50x microscope objective lens to a spot size of ~2 μm. The backscattered signal was collected through the same objective lens and dispersed by a 2400 g/mm grating with a spectral resolution of ~0.4 cm$^{-1}$. The laser power was kept below 0.25 mW to avoid sample damage and laser heating.

**Acknowledgements**

We thank Qihua Xiong, Bo Sun, Lexian Yang, and Yong Xu for helpful discussions. Sample preparation and ultrafast optical measurements were carried out at Tsinghua University. Raman studies were performed at the Institute of Semiconductors, CAS. The work was supported by the National Key R&D Program of China (Grant Nos. 2020YFA0308800 and 2021YFA1400100), the National Natural Science Foundation of China (Grant Nos. 12074212, 12004377, 11874350, 51991343 and 21975140), CAS Key Research Program of Frontier Sciences (Grant Nos. ZDBS-LY-SLH004 and XDPB22) and the Beijing Advanced Innovation Center for Future Chip (ICFC).

# Ultrafast coherent interlayer phonon dynamics in atomically thin layers of MnBi$_2$Te$_4$

F. Michael Bartram[1,2], Yu-Chen Leng[3], Yongchao Wang[4], Liangyang Liu[1], Xue Chen[3], Huining Peng[1], Hao Li[7], Pu Yu[1,5,6], Yang Wu[7,8], Miao-Ling Lin[3], Jinsong Zhang[1,5], Ping-Heng Tan[3] and Luyi Yang[1,5,]*

[1]State Key Laboratory of Low Dimensional Quantum Physics, Department of Physics, Tsinghua University, Beijing 100084, China

[2]Department of Physics, University of Toronto, Toronto, Ontario M5S 1A7, Canada

[3]State Key Laboratory of Superlattices and Microstructures, Institute of Semiconductors, Chinese Academy of Sciences, Beijing 100083, China

[4]Beijing Innovation Center for Future Chips, Tsinghua University, Beijing, China

[5]Frontier Science Center for Quantum Information, Beijing 100084, China

[6]RIKEN Center for Emergent Matter Science (CEMS), Wako 351-198, Japan

[7]Tsinghua-Foxconn Nanotechnology Research Center, Department of Physics, Tsinghua University, Beijing 100084, China

[8]Beijing Univ Chem Technol, Coll Sci, Beijing, 100029, China

*Email: luyi-yang@mail.tsinghua.edu.cn

## Table of Contents

# 1. Symmetry analysis of the Raman selection rules and polarization dependence of Raman spectra

The primitive unit cell of MBT consists of one Mn, two Bi and four Te atoms, for a total of 7 atoms. As a result, there are 21 phonon modes in a bulk crystal, of which 3 are acoustic modes and 18 are optical modes. In the paramagnetic phase, bulk MBT belongs to the space group $R\bar{3}m$(166) with point group $D_{3d}$. The irreducible representations of the phonon modes at $\Gamma$ point can be written as $\Gamma_{bulk}$ = 3$A_{1g}$ + 4$A_{2u}$ + 3$E_g$ + 4$E_u$, among which 9 modes are Raman active (3 with $A_{1g}$ representation and 3 with doubly degenerate $E_g$ representation) and 12 IR active modes including the acoustic modes. The A and E symmetry modes represents out-of-plane (breathing) and in-plane (shear) vibrations, respectively.

The Raman tensors are given by

$$R(A_{1g}) = \begin{pmatrix} a & 0 & 0 \\ 0 & a & 0 \\ 0 & 0 & b \end{pmatrix}, \qquad (S1)$$

$$R^{(a)}(E_g) = \begin{pmatrix} c & 0 & 0 \\ 0 & -c & d \\ 0 & d & 0 \end{pmatrix} \text{ and } R^{(b)}(E_g) = \begin{pmatrix} 0 & -c & -d \\ -c & 0 & 0 \\ -d & 0 & 0 \end{pmatrix}. \qquad (S2)$$

The Raman intensity of a phonon mode is given by $I \propto |e_i \cdot R \cdot e_s|^2$, where $e_i$ and $e_s$ are the polarizations of the incident and scattered light, respectively. For polarized Raman measurements in a backscattering geometry, symmetry selection rules predict that both $A_{1g}$ and $E_g$ modes can be observed in the parallel polarization geometry, while only the $E_g$ mode is observable under the perpendicular polarization configuration due to the nonzero off-diagonal Raman tensor elements.

In few layer MBT 2D crystals, the crystal symmetry is lowered with respect to that of its bulk counterpart due to the lack of the translation symmetry along the c axis. Few-layer ABC stacking MBT in the paramagnetic phase belongs to the space group $P\bar{3}m1$(164) with the same point group $D_{3d}$ as the bulk. The primitive unit cell of an $N$SL crystal is composed of 7$N$ atoms, and the irreducible representation of the Brillouin zone center phonons can be written as $\Gamma_{odd} = \frac{7N-1}{2}(3A_{1g} + 3E_g) + \frac{7N+1}{2}(4A_{2u} + 4E_u)$, $N$ = 1, 3, 5, … for odd-SL layers, and $\Gamma_{\text{even}} = \frac{7N}{2}(3A_{1g} + 3E_g + 4A_{2u} + 4E_u)$, $N$ = 2,4, 6, … for even-SL layers. Similar to the bulk case, all the $A_{1g}$ and $E_g$ modes are Raman active in few-SL MBT. Both the Raman tensors and selection rules are the same as that in the bulk case. However, contrary to the bulk case the interlayer vibration modes are allowed [S1].

In Figure S1, we show the full polarized Raman spectra for one of the sample regions. We assign the vibration modes according to the Raman selection rules.

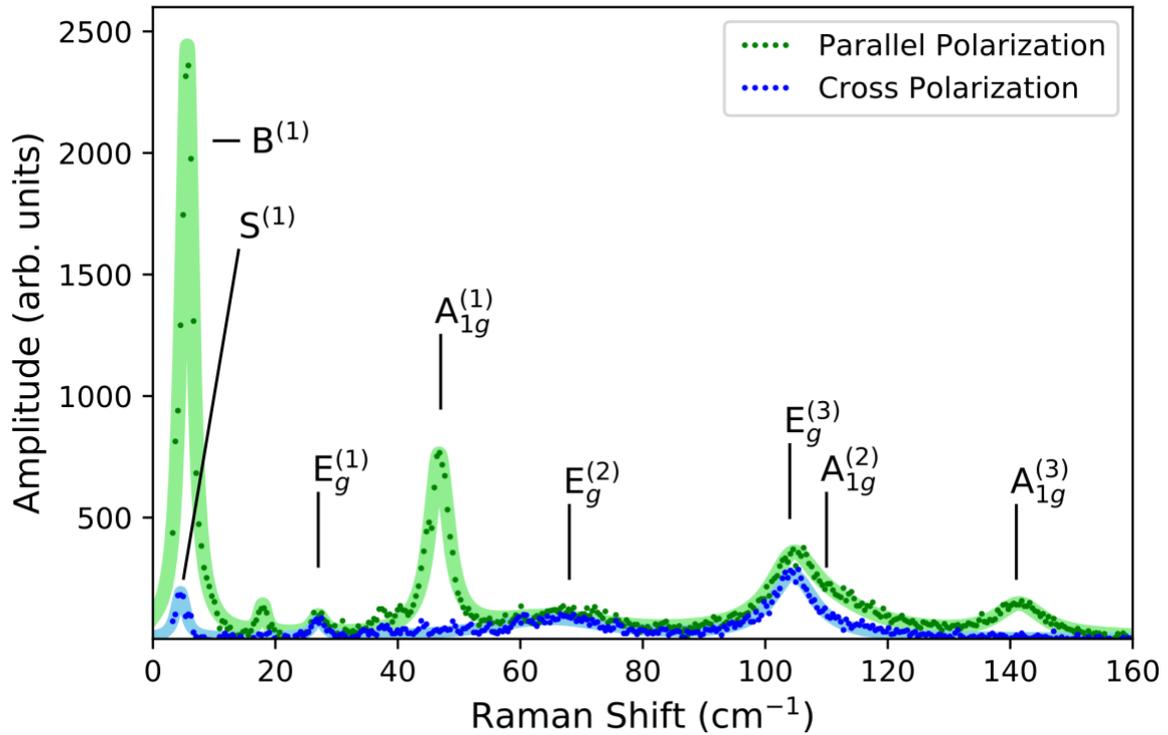

*Figure S1 Polarized Raman spectroscopy for a 7-layer region of MBT. The $A_{1g}$ and B peaks both vanish in a cross-polarized measurement.*

The irreducible representations for a monolayer MBT are the same as the bulk MBT, since both 1SL and bulk MBT have the same numbers of atoms in a primitive cell. Going from 1SL to $N$SL, each of the 21 normal modes in 1SL will evolve into $N$ modes in $N$SL, maintaining the same vibration mode within each SL, while varying the relative displacement phases among different SLs. The out-of-plane/in-plane acoustic mode in 1SL (bulk) splits into one out-of-plane/in-plane acoustic mode and $N$-1 interlayer breathing/shear modes in $N$SL. Among the interlayer breathing and shear modes, the $(2j-1)^{\text{th}}$ lowest frequency modes are Raman active ($A_{1g}$ and $E_g$) and the $(2j)^{th}$ lowest frequency modes are IR active ($A_{2u}$ and $E_u$), where $j$ = 1, 2, 3… This is similar to $Bi_2Te_3$ and $Bi_2Se_3$ [S5], which have the same symmetry as MBT.

## 2. Additional Raman data

In Figure S2a, we have plotted the change in location of the $A_{1g}^{(1)}$ peak as a function of inverse layer number, which redshifts for thinner sample regions, similar to previous reports [S2]. We also plot the linewidths obtained from fits to the Raman spectra for the $B^{(1)}$ and $S^{(1)}$ peaks in Figure S2b for comparison with the decay rates from the time-resolved data (where the linewidth for a signal decaying as $e^{-\gamma t}$ is $\gamma/2\pi$). The Raman linewidths are quite noisy and in thinner regions are noticeably larger than the time-resolved decay rates. This is probably attributable to inaccuracies from the background, as well as the fact that these linewidths are relatively small compared to the frequency resolution of the instrument.

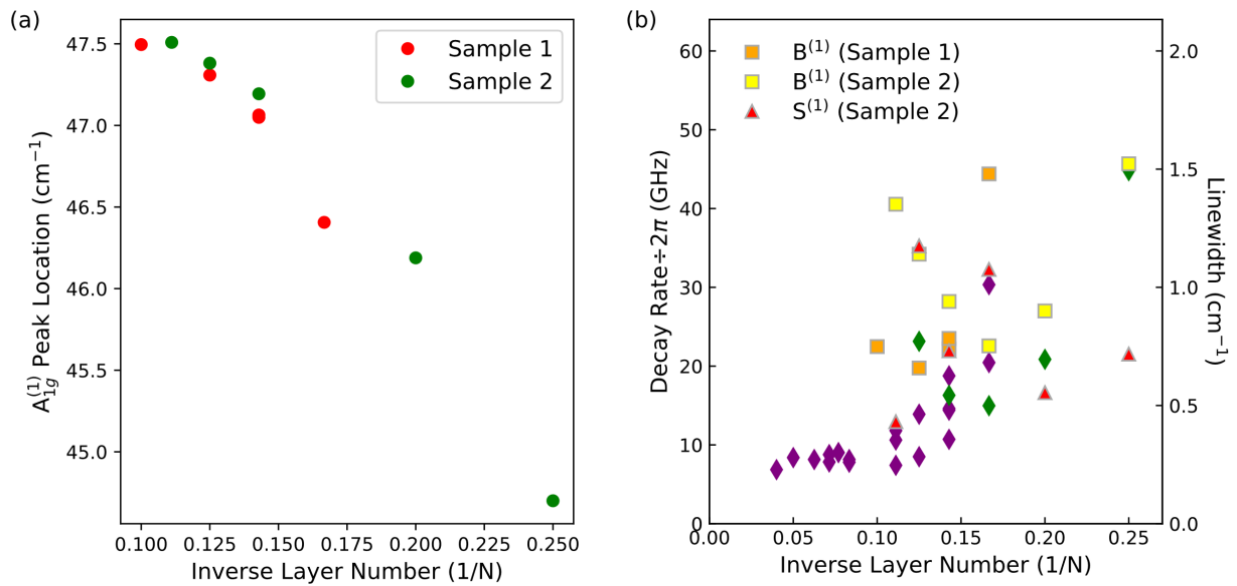

*Figure S2 Additional data from Raman measurements. The thickness dependence of the $A_{1g}^{(1)}$ peak (a) shows a clear redshift as the sample becomes thinner. We also show the linewidths of the $B^{(1)}$ and $S^{(1)}$ peaks (c) alongside the decay rates observed in the time-resolved data divided by $2\pi$ (purple and green diamond symbols for Samples 1 and 2 respectively) for comparison.*

## 3. Polarization dependence of the time-resolved reflectivity

In our experiment both the pump and probe beams are at normal incidence (propagate along the c axis). One can show that coherent phonon signal of the symmetric $A_{1g}$ mode is independent of θ, while the oscillatory signal of the $E_g$ mode should vary as cos(2θ), where θ is the angle between the polarizations of the pump and probe beams [S3,S4]. In Fig. S3, we show that the transient reflection signals are basically identical for θ = 0, or π/2, revealing that the coherent phonon mode is an $A_{1g}$ mode.

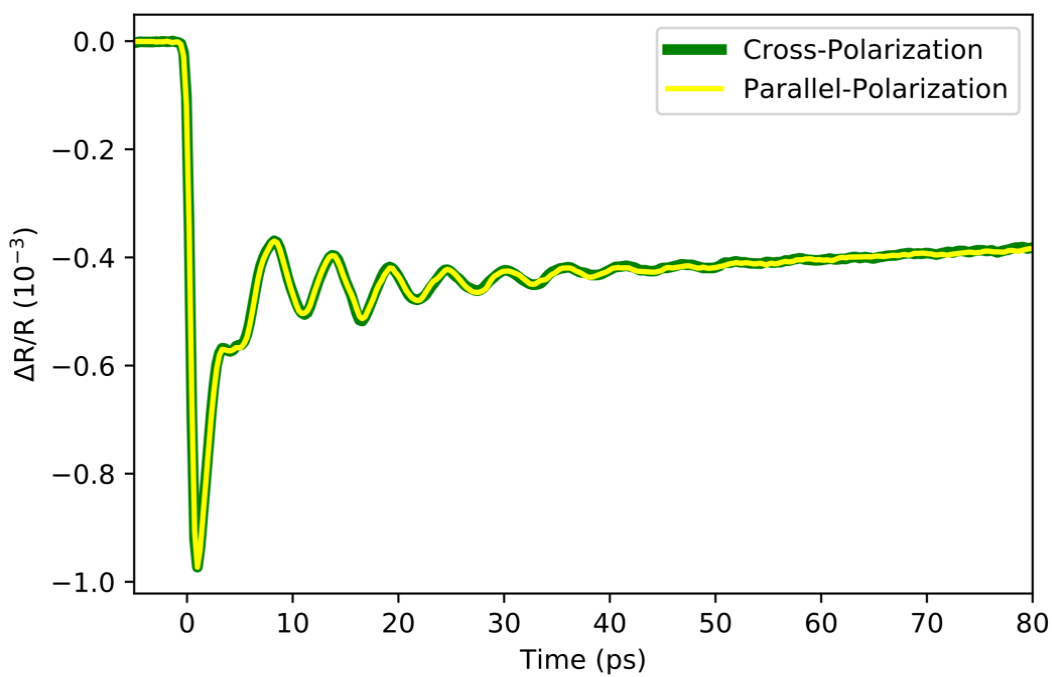

*Figure S3 Time-resolved reflectivity changes when the pump is polarized parallel (thin yellow line) or perpendicular (thick green line) to the probe beam, for a 7-layer region of MBT. The data in both cases is essentially identical.*

## 4. Wavelength dependence of the time-resolved reflectivity

The wavelength of light used for the pump and probe beams will generally influence the observed reflectivity changes. In coherent phonon measurements, wavelength dependence can be an important tool to distinguish standing waves in ultrathin samples from traveling acoustic phonons. In the former case, the wavelength of light is not relevant, but in the latter the observed frequency will be proportional to the optical frequency of the probe beam [S9]. Figure S4 shows our results using various pump/probe wavelength configurations, with the observed phonon frequencies showing a clear lack of any significant wavelength dependence.

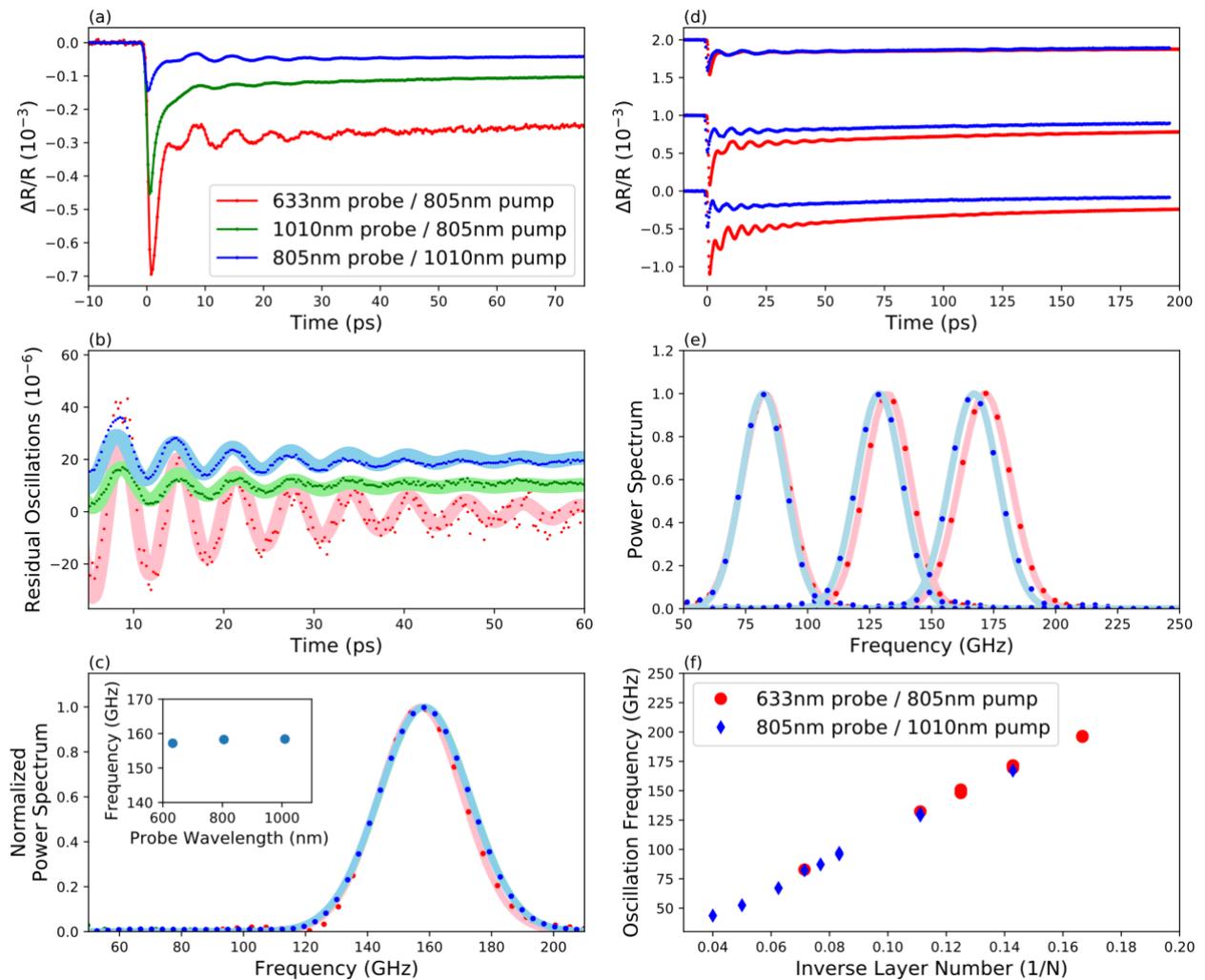

*Figure S4 Time-resolved reflectivity measurements for different laser wavelength configurations. In panels (a)-(c), we contrast three measurements done with varying pump-probe combinations for a single location, showing how although the signal changes somewhat between the three, the basic shape, and more importantly the frequency of the observed oscillation remains consistent. In (d)-(e), we compare the two datasets for the same sample used in figure 3 in the main text. (d) and (e) show raw data and normalized power spectra respectively for the three overlapping points, and (f) shows which points in the main figure 3 plot correspond to which setup.*

# 5. Field and temperature dependence of the time-resolved reflectivity

We also performed measurements at low temperature, where MBT becomes magnetic, to determine how this affected the interlayer phonons. Figures S5 and S6 show measurements of the interlayer phonons on a 6-layer sample, across temperature and field respectively. As the temperature is decreased, we observe a small but clear increase in the phonon frequency, corresponding to a roughly 10% increase in the coupling strength at 3 K compared to room temperature. However, across the transition temperature at ~20 K, there is no visible change in this trend, suggesting that interactions between the magnetic order and the interlayer phonons are relatively small. We verified this by applying fields of up to 6 T in the out-of-plane direction while the sample was cooled to 3 K, where we find no significant changes in the phonon frequency or decay rate as the field is varied.

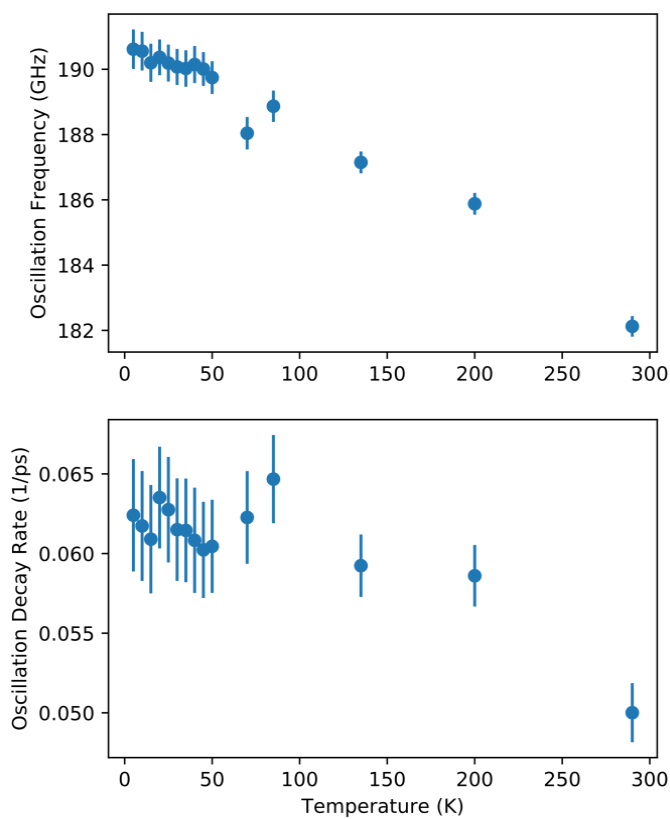

*Figure S5 Results of time-resolved reflectivity measurements for varying sample temperature. A small but noticeable shift in frequency occurs when the sample cools down. However, there is no visible shift when dropping below $T_C$ (~20K). The decay rate of the phonons also is roughly constant across the temperature range, with the occasional shifts (notable a small drop at room temperature) likely due to slight sample motion as the temperature is increased.*

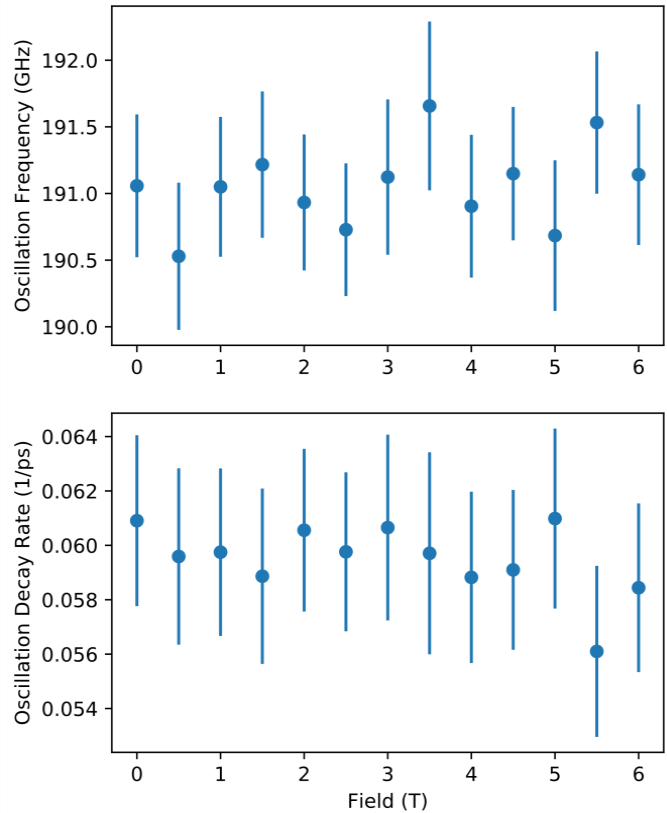

*Figure S6 Results of time-resolved reflectivity measurements for applied external fields. Both the frequency and the decay rate are very stable up to an applied field of 6T.*

## 6. Pump fluence dependence of the time-resolved reflectivity

Figure S7 shows measurements done on an 8-layer region of the sample at room temperature with varying pump power. Looking at the signal after being scaled by the pump power, it is obvious that within the first few picoseconds we see some deviation from linearity, indicating that the short-lived component may be starting to saturate. At later times however, the rescaled signal appears close to identical. This is further demonstrated by plotting the amplitude (c) and lifetime (d) of the oscillatory component against signal power, where the amplitude shows excellent linear scaling, and the lifetime does not show any power dependence. Note that the frequency does shift very slightly, as seen in the power spectra (b), most likely due to sample heating at higher pump power combined with the previously mentioned temperature dependence of the frequency.

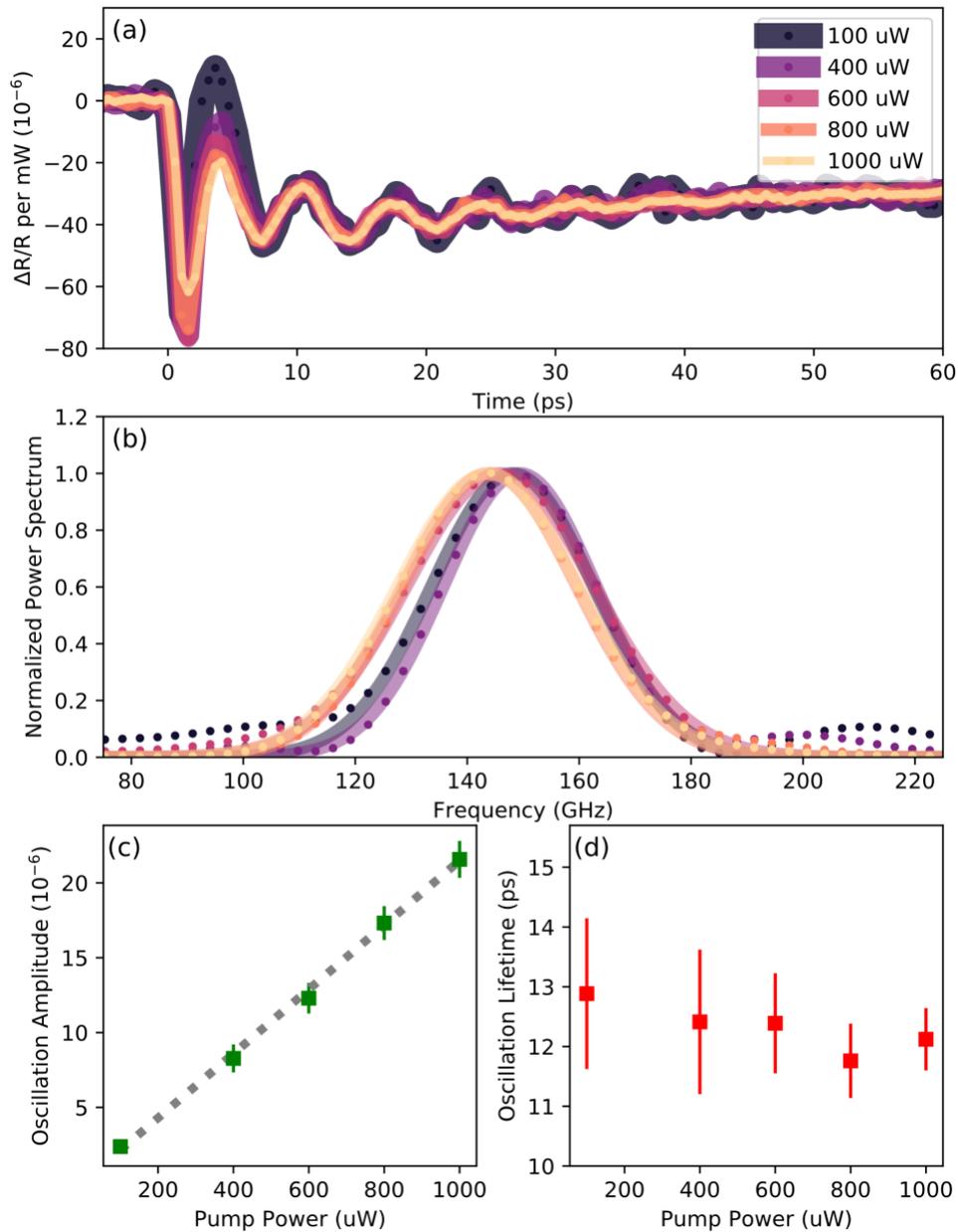

*Figure S7 Time-resolved reflectivity measurements with a pump beam varying in power. The raw data, normalized to the pump power is shown in (a), and the normalized power spectra in (b). The amplitude of the oscillation (c) scales linearly with the power, and the lifetime (d) remains constant.*

# 7. Additional time-resolved reflectivity measurements

In Figure 3, we showed the results of time-resolved reflectivity measurements on a variety of sample regions. In addition to the data already shown in Figure 2, we obtained results from some even thinner areas on a second sample, a few of which are shown in Figure S8. In addition, in Figure S9 we plot a few of the much thicker regions of our first sample (up to 25 layers thick), which were not shown in the main text.

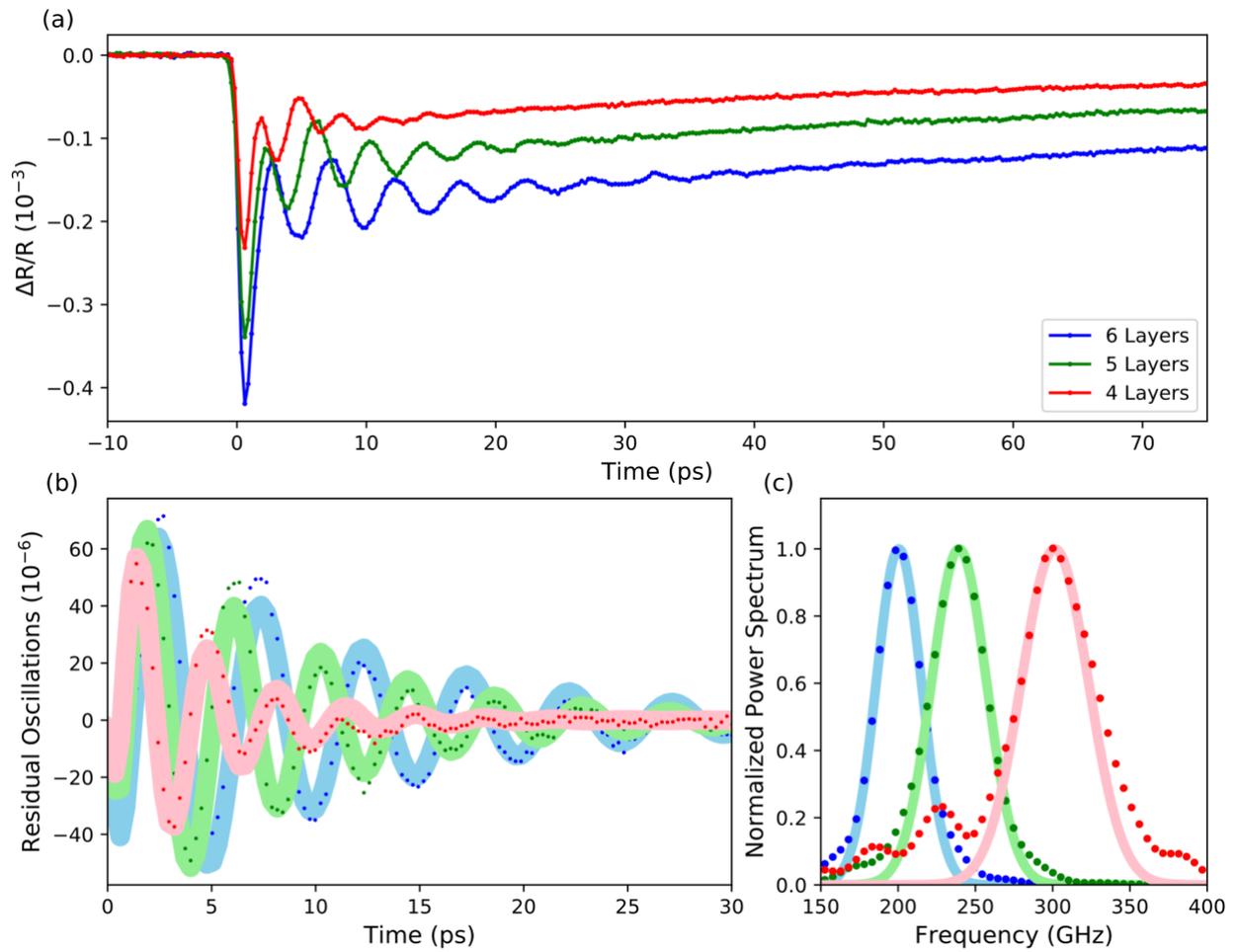

Figure S8 Time-resolved reflectivity measurements, showing raw data (a), the oscillatory component (b), and the FFT of the oscillatory component (c) for selected regions on Sample 2.

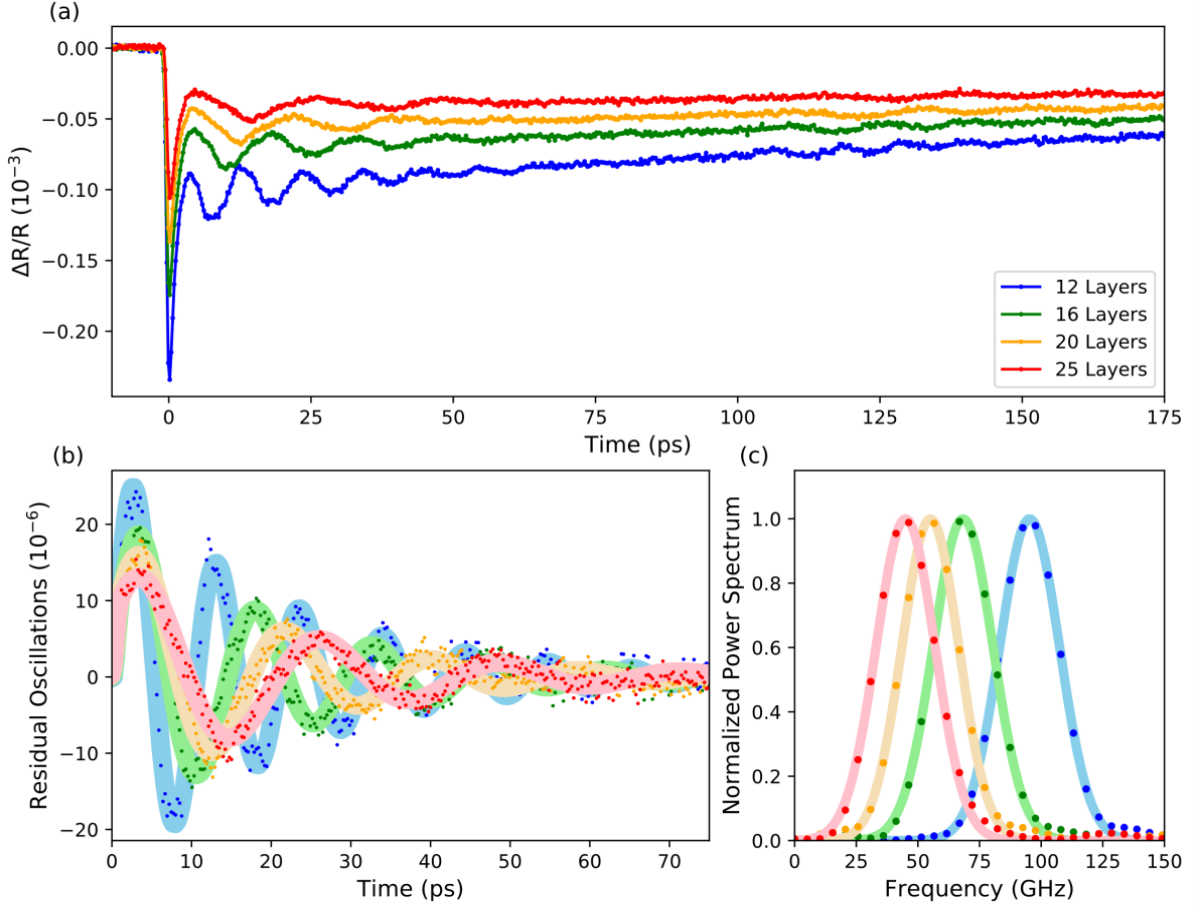

*Figure S9 Time-resolved reflectivity measurements, showing raw data (a), the oscillatory component (b), and the FFT of the oscillatory component (c) for some of the thicker regions on Sample 1.*

## 8. Comparison of the out-of-plane and in-plane interlayer mechanical parameters

We compare the mechanical parameters of MBT with a few selected 2D materials. The characteristic frequencies $f_{0,b}$ and $f_{0,s}$ of the interlayer breathing/shear mode (LBM/LSM) force constants were obtained from the fits of the frequency vs layer number to the linear chain model. The LBM/LSM force constant per unit area are given by $k_z = \mu(\pi f_{0,b})^2$ and $k_x = \mu(\pi f_{0,s})^2$. The LBM and LSM force constants per effective atom $k_{z/atom}$ and $k_{x/atom}$ were calculated from $k_z$ and $k_x$ accounting the number of bottom atoms per unit cell. The longitudinal/transverse acoustic velocity along the out-of-plane direction is given by $v_{B/S} =$ $d\sqrt{\frac{k_{z/x}}{\mu}}$. The elastic constants $C_{33} = k_z d = \rho v_B^2$ and $C_{44} = k_x d = \rho v_S^2$. The acoustic impedance is given by $Z = \rho v_B$.

Table S1 Comparison of the out-of-plane and in-plane interlayer mechanical parameters with selected 2D layered materials.

| | 2D material | MnBi$_2$Te$_4$ | Bi$_2$Se$_3$ | Bi$_2$Te$_3$ | Graphene | MoS$_2$ |
|---|---|---|---|---|---|---|
| | Lattice const. $a$ (Å) | 4.33 | 4.13 | 4.39 | 2.46 | 3.16 |
| | Interlayer distance $d$ (Å) | 13.6 | 9.53 | 10.2 | 3.35 | 6.15 |
| | Point group | D$_{3d}$ | D$_{3d}$ | D$_{3d}$ | D$_{6h}$ | D$_{6h}$ |
| | $\rho$ (10$^3$ kg m$^{-3}$) | 7.36 | 7.71 | 7.84 | 2.28 | 5.00 |
| | $\mu$ (10$^{-6}$ kg m$^{-2}$) | 10.0 | 7.35 | 8.00 | 0.76 | 3.08 |
| LBM | $k_z$ (10$^{19}$ N m$^{-3}$) | 5.7 | 5.26 | 13.33 | 11.8 | 8.62 |
| | $k_{z/atom}$ (N m$^{-1}$) | 9.3 | 7.8 | 22.2 | 3.1 | 7.5 |
| | $C_{33}$ (GPa) | 77.5 | 50.1 | 136.0 | 39.5 | 53.0 |
| | $v_B$ (km s$^{-1}$) | 3.25 | 2.55 | 4.16 | 4.16 | 3.27 |
| | $f_{0,B}$ (THz) | 0.76 | 0.85 | 1.30 | 3.96 | 1.69 |
| LSM | $k_x$ (10$^{19}$ N m$^{-3}$) | 2.9 | 2.27 | 4.57 | 1.28 | 2.72 |
| | $k_{x/atom}$ (N m$^{-1}$) | 4.7 | 3.4 | 7.6 | 0.34 | 2.4 |
| | $C_{44}$ (GPa) | 39.4 | 21.6 | 46.6 | 4.29 | 16.7 |
| | $v_S$ (km s$^{-1}$) | 2.31 | 1.68 | 2.44 | 1.37 | 1.83 |
| | $f_{0,S}$ (THz) | 0.55 | 0.56 | 0.76 | 1.30 | 0.95 |
| | $Z$ (MPa s m$^{-1}$) | 23.9 | 19.7 | 32.6 | 9.5 | 16.3 |
| | Reference | This work | [S5] | [S5] | [S6,S7] | [S8] |

## 9. The linear chain model

The linear chain model is a simple model for understanding vibrations in a layered system. Each layer is treated as moving as a single unit, with a mass per unit area given by $\mu$, and is coupled to neighboring layers with a spring constant per unit area $K$. When the number of layers $N \to \infty$, this gives rise to interlayer acoustic phonons, where the phonon velocity will be given by $v = d\sqrt{k/\mu}$, with $d$ the spacing between layers. For a small number of layers, however, standing waves are formed, with the frequencies of those standing waves strongly dependent on the number of layers. These frequencies may, however, be complicated by the interface with the substrate. The model discussed in our paper is one of the simplest, where the substrate is ignored, and the sample is treated as if it were free-standing, giving frequencies

$$f_{n,N} = \frac{1}{\pi}\sqrt{\frac{k}{\mu}}\sin\left(\frac{\pi n}{2N}\right)$$

where the mode number $n$ runs from 1 to $N - 1$. More realistically, this can be viewed as a limiting case of a vanishingly weak substrate interface coupling $k_i \to 0$ (where the interface position is fixed), with a low frequency "interface mode" (where the sample moves as a single unit) given by frequency $f = \frac{1}{2\pi}\sqrt{k_i/N\mu}$. We note that in practice, reaching a regime where the effect of the substrate is negligible is possible even with a relatively large $k_i$ if the substrate is soft and light enough that it can move along with the bottom layer without significantly impeding its motion. Quantitatively, this can be described by the impedance $Z = \rho v$ being much lower in the substrate than in the sample. A good example of this can be seen in reference [S5], in which the interface mode was observed in 2- to 4-layer Bi$_2$Te$_3$ on two different substrates, SiO$_2$/Si and mica, where the interface mode changed frequency due to the variation in coupling strength, but the interlayer modes did not, suggesting negligible influence of the interfacial coupling strength on the interlayer mode frequencies. Given the similarity of Bi$_2$Te$_3$ to MBT, it is very likely that a similar case holds for our results. Calculating the impedances using values from Table 1 gives further credence to this conclusion, with values of roughly 33 and 24 MPa s m$^{-1}$ for Bi$_2$Te$_3$ and MBT respectively, compared to a value of 13 MPa s m$^{-1}$ (using $v = 6$ km s$^{-1}$, $\rho$ = 2.2 Mg m$^{-3}$) for SiO$_2$.

We also note that the at the opposite extreme, we can imagine the substrate to be perfectly rigid and very tightly bound to the sample ($k_i \to \infty$), rendering the bottom layer immobile. In this case, the frequencies would be described by

$$f_{n,N} = \frac{1}{\pi}\sqrt{\frac{k}{\mu}}\sin\left(\frac{\pi(2n-1)}{4N-2}\right)$$

with again the mode number running from 1 to *N*-1. Note that for large *N*, the frequency of the lowest-order mode is $\approx 1/4N\sqrt{k/\mu}$, compared to $\approx 1/2N\sqrt{k/\mu}$ for the previous case. This factor-of-two difference is intuitively obvious when considering the equivalent problem in the continuous limit, where for open-closed boundary conditions the sample thickness is equal to ¼ of a wavelength, compared to open-open boundary conditions where the sample thickness is ½ of a wavelength. Because the *N*-dependence in both cases is very similar, it is difficult to conclusively rule out this situation, which would lead to large differences in the extracted force constants and phonon velocities (by factors of roughly 4 and 2 respectively). However, this would also result in an acoustic impedance that is 2 times larger (~48 MPa s m$^{-1}$), making it significantly larger than that of SiO$_2$ and contradicting the initial assumption of the substrate being relatively rigid (which requires $Z_{MBT} < Z_{SiO2}$). Given the above discussion, along with the fact that the obtained force constants are in line with similar layer materials, the free-standing model seems to be the better choice.

We notice that a very pronounced peak due to the interface coupling is observed by Raman spectroscopy in a very recent publication [S9]. By contrast, the lowest breathing mode is very weak in stark contrast to our case. This may be due to different sample fabrication procedures, resulting in different coupling strength to the substrate. Nevertheless, the exacted breathing mode force constant is $k_z = 4.3 - 6.4 * 10^{19}$ N/m$^3$, in a good agreement with our results. The variation in $k_z$ depends on the modeling of the interface coupling $k_i$.